\documentstyle[12pt,epsfig,axodraw,a4]{article} 
\newcommand{\vs}{\vspace{-0.25cm}}
\begin{document} 
\begin{center}
{\Large{\bf Chiral dynamics of $\Sigma$-hyperons in the nuclear medium}}  
\bigskip

N. Kaiser\\
\medskip
{\small Physik-Department T39, Technische Universit\"{a}t M\"{u}nchen,
    D-85747 Garching, Germany}
\end{center}
\medskip
\begin{abstract}
Using SU(3) chiral perturbation theory we calculate the density-dependent
complex mean field $U_\Sigma(k_f)+ i\, W_\Sigma(k_f)$ of a $\Sigma$-hyperon in
isospin-symmetric nuclear matter. The leading long-range $\Sigma N
$-interaction arises from one-kaon exchange and from two-pion exchange with a 
$\Sigma$- or a $\Lambda$-hyperon in the intermediate state. We find from the 
$\Sigma N\to \Lambda N$ conversion process at nuclear matter saturation
density $\rho_0 = 0.16\,$fm$^{-3}$ an imaginary single-particle potential of
$W_\Sigma(k_{f0}) =-21.5$\,MeV, in fair agreement with existing empirical 
determinations. The genuine long-range contributions from iterated (second
order) one-pion exchange with an intermediate $\Sigma$- or $\Lambda$-hyperon
sum up to a moderately repulsive real single-particle potential of
$U_\Sigma(k_{f0})= 59$\,MeV. Recently measured $(\pi^-,K^+$) inclusive spectra
related to $\Sigma^-$-formation in heavy nuclei give evidence for a
$\Sigma$-nucleus repulsion of similar size. Our results suggest that the net
effect of the short-range $\Sigma N$-interaction on the $\Sigma$-nuclear mean
field could be small.
\end{abstract}

\bigskip


\vspace{0.5cm}

The $\Sigma$-nucleus optical potential describes the behavior of a 
$\Sigma$-hyperon in the nuclear medium. The quantitative determination of this
(complex) potential is a subject of current interest. Whereas an earlier 
analysis of the shifts and widths of X-ray transitions in $\Sigma^-$-atoms has
come up with an attractive (real) $\Sigma$-nucleus potential of about $-27\,
$MeV \cite{dover} (i.e. almost equal to well-established attractive 
$\Lambda$-nucleus potential of depth $-28\,$MeV \cite{chrien}) there is 
nowadays good experimental and phenomenological evidence for a substantial 
$\Sigma$-nucleus repulsion. A reanalysis of the $\Sigma^-$-atom data by Batty, 
Friedman and Gal \cite{batty} including the then available precise 
measurements of W and Pb atoms and employing phenomenological 
density-dependent fits has lead to a $\Sigma$-nucleus potential with a 
strongly repulsive core (of height $\sim 95$\,MeV) and a shallow attractive 
tail outside the nucleus. However, because of the poor penetration of the 
$\Sigma^-$-hyperon into the nucleus such fitted potentials are not well 
defined by the $\Sigma^-$-atom data in the nuclear interior. The inclusive 
($\pi^-,K^+$) spectra on medium-to-heavy nuclear targets measured at KEK 
\cite{noumi,saha} give a more direct evidence for a strongly repulsive 
$\Sigma$-nucleus potential. In the framework of the distorted-wave impulse 
approximation a best fit of the measured ($\pi^-,K^+$) inclusive spectra on 
Si, Ni, In and Bi targets is obtained with a $\Sigma$-nucleus repulsion of 
about $90\,$MeV \cite{saha}. In addition a nonzero value of the imaginary 
$\Sigma$-nucleus potential (with a best fit value of about $-40\,$MeV 
\cite{saha}) is also required in order to reproduce the observed spectra of 
the double differential cross section $d^2\sigma/d \Omega dE$. Very recently, 
Kohno et al. \cite{kohno} have calculated the ($\pi^-,K^+$) inclusive spectra 
on Si within a semiclassical distorted wave model and they found that the KEK 
data can also be well reproduced with a complex $\Sigma$-nucleus potential of 
strength $(30-20\,i)\,$MeV. A possible reason for the different result from 
ref.\cite{saha} is due to avoiding the factorization approximation by an 
average cross section \cite{kohno}. For an up-to-date and comprehensive 
overview of hypernuclear physics see also ref.\cite{galneu}.    

In the standard one-boson exchange models for hyperon-nucleon interaction
there are appreciable uncertainties in various meson-baryon coupling 
constants, although SU(3) relations are imposed. Most of these models give an 
attractive (real) $\Sigma$-nucleus potential \cite{rolf,schulze,yamamoto}, but
the Nijmegen model F \cite{nijmf} leads to repulsion, estimated to be about 
$(4 - 8\,i)\,$MeV in nuclear matter \cite{yamada}. A nonrelativistic SU(6) 
quark model for the unifying description of octet baryon-baryon interactions 
has been developed by the Kyoto-Niigata group \cite{fuji}. G-matrix 
calculations in lowest order Brueckner theory \cite{kohno2} with this 
hyperon-nucleon interaction showed that the real part of the $\Sigma$-nuclear 
mean-field is repulsive of the order of $20\,$MeV due to a strong repulsion in 
the total $\Sigma N$-isospin $3/2$ channel which originates from Pauli 
blocking effects at the quark level. The same calculation \cite{kohno2} has 
found an imaginary part of the $\Sigma$ single-particle potential in nuclear 
matter of $-20\,$MeV, comparable to the value $-16\,$MeV extracted in the 
earlier analysis of the $\Sigma^-$-atom data \cite{dover}. The basic physical 
mechanism behind this sizeable (negative) imaginary part is of course the 
strong conversion process $\Sigma N \to \Lambda N$ in nuclear matter.  
             
More recently, chiral effective field theory approaches have opened new 
perspectives for dealing with binding and saturation of nuclear matter as well
as single-particle properties of nucleons \cite{nucmat} and $\Lambda$-hyperons 
\cite{lambdapot} in the nuclear medium. A key element in these approaches is
the separation of long- and short-distance dynamics and an ordering scheme in
powers of small momenta. At nuclear matter saturation density $\rho_0=0.16\,
$fm$^{-3}$ the Fermi momentum $k_{f0}$ and the pion mass $m_\pi$ are
comparable scales ($k_{f0} \simeq 2m_\pi$) and therefore pions must be included
as explicit degrees of freedom in the description of the nuclear many-body
dynamics. In this work we extend such a field-theoretical approach to the 
complex single-particle potential of a $\Sigma$-hyperon in isospin-symmetric 
nuclear matter. 

Our calculation is based on the leading order chiral meson-baryon Lagrangian
in flavor SU(3) \cite{review}: 
\begin{equation} {\cal L}_{\phi B} = {D\over 2f_\pi} \, {\rm tr}(\bar B\,
\vec \sigma \cdot \{\vec \nabla \phi, B\})+{F\over 2f_\pi} \, {\rm tr}(\bar 
B\,\vec \sigma \cdot [\vec \nabla \phi,B]) +\dots \,, \end{equation}
where the traceless hermitian $3\times 3$ matrices $B$ and $\phi$ collect
the octet baryon fields ($N,\Lambda,\Sigma,\Xi$) and the pseudoscalar 
Goldstone bosons fields ($\pi,K,\bar K,\eta$), respectively. The parameter 
$f_\pi = 92.4\,$MeV is the weak pion decay constant and $D$ and $F$ denote the
SU(3) axial-vector coupling constants of the octet baryons. We choose as their
values $D= 0.84$ and $F=0.46$. This leads to a $KN\Sigma$-coupling constant of
$g_{KN\Sigma}=(D-F)(M_N+M_\Sigma)/(2f_\pi)=4.4$, a $\pi\Lambda\Sigma$-coupling 
constant of $g_{\pi\Lambda\Sigma} = D(M_\Lambda+M_\Sigma)/(\sqrt{3}f_\pi)=
12.1$ and a $\pi\Sigma \Sigma $-coupling constant of $g_{\pi\Sigma\Sigma}= 2F 
M_\Sigma/f_\pi = 11.9$, consistent with the empirical values summarized in 
Tables 6.3 and 6.4 of ref.\cite{coupl}. The $\pi\Lambda\Sigma$-coupling 
constant used in the present work is also consistent with the value $g_{\pi
\Lambda\Sigma}= 12.9\pm 0.9$ extracted recently from hyperonic atom data in 
ref.\cite{lois}. Furthermore, the pion-nucleon coupling constant has the value
$g_{\pi N}= g_A  M_N/f_\pi= 13.2$ \cite{pavan} with $g_A =D+F=1.3$. The
ellipsis in eq.(1) stands for the chiral-invariant interaction terms with two 
and more Goldstone boson fields which do not come into play in the present 
calculation.

\begin{figure}
\begin{center}
\includegraphics[scale=1.15]{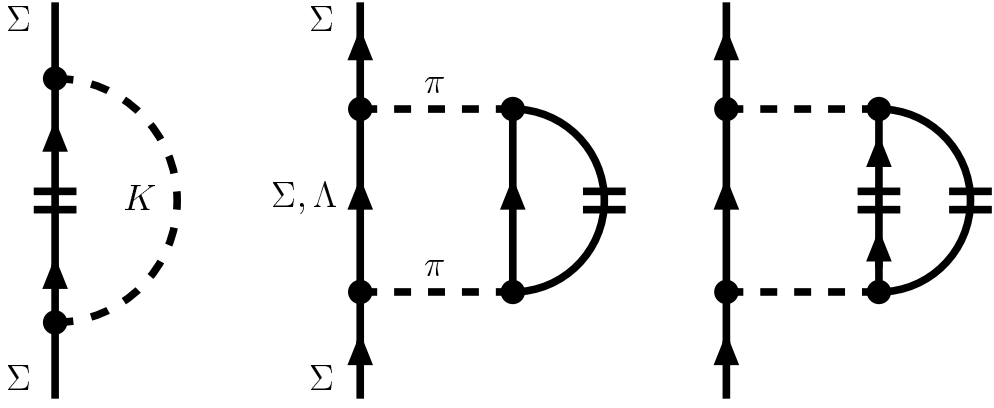}
\end{center}
\vspace{-.4cm}
\caption{One-kaon exchange Fock diagram and iterated one-pion exchange Hartree
diagrams with $\Sigma$- or $\Lambda$-hyperons in the intermediate state. The
horizontal double-line symbolizes the filled Fermi sea of nucleons, i.e. the
medium insertion $-\theta(k_f-|\vec p\,|)$ in the in-medium nucleon propagator
\cite{nucmat}. Effectively, the medium insertion sums up hole-propagation and 
the absence of particle-propagation below the Fermi surface $|\vec p\,|< k_f$.}
\end{figure}

Consider the density-dependent complex selfenergy $U_\Sigma(k_f)+i\,W_\Sigma
(k_f)$ of a zero-momentum $\Sigma$-hyperon ($\vec p_\Sigma = \vec 0$) placed as
a test particle into isospin-symmetric nuclear matter. The value $U_\Sigma(
k_{f0})+i\,W_\Sigma( k_{f0})$ at nuclear matter saturation density $\rho_0 = 
0.16\,$fm$^{-3}$ fixes the strength of the $\Sigma$-nucleus optical potential.
We calculate the long-range contributions to the $\Sigma$-nuclear mean field
$U_\Sigma(k_f)+i\,W_\Sigma(k_f)$ generated by the exchange of light Goldstone
bosons between the $\Sigma$-hyperon and the nucleons in the filled Fermi sea.
The only nonvanishing one-meson exchange contribution comes from the
kaon-exchange Fock diagram in Fig.\,1, from which we obtain the following
(small and) repulsive contribution to the real part of the $\Sigma$-nuclear
mean field:  
\begin{equation} U_\Sigma(k_f)^{(K)} = {(D-F)^2 \over (2\pi f_\pi)^2} \Bigg\{
{k_f^3\over 3} -  m_K^2  k_f+ m_K^3 \arctan{k_f \over m_K} \Bigg\}\,, 
\end{equation}
with $m_K = 496\,$MeV the average kaon mass. At densities at and below
nuclear matter saturation density $\rho \leq 0.16\,$fm$^{-3}$ (corresponding
to Fermi momenta $k_f \leq 263\,$MeV) the kaon-exchange can already be 
regarded as being of short range. The ratio $k_f/m_K \leq 0.53$ is small 
and the expression in curly brackets of eq.(2) is dominated by its leading 
term $k_f^5/5m_K^2$ in the $k_f$-expansion. 

The truly long-range interaction between the $\Sigma$-hyperon and the 
nucleons arises therefore from two-pion exchange. The corresponding two-loop 
diagrams with a $\Sigma$- or a $\Lambda$-hyperon in the intermediate state are 
shown in Fig.\,1. Their relative isospin factor (for the $\Sigma$- and 
$\Lambda$-intermediate state) is $6(F/D)^2 \simeq 1.8$. We find from the 
second diagram in Fig.\,1 with one medium insertion \cite{nucmat} and a 
$\Sigma$-hyperon in the intermediate state the following long-range 
contribution to the real part of the $\Sigma$-nuclear mean-field:\footnote{We 
have used dimensional regularization (where a linear divergence $\int_0^\infty 
dl\,1$ is set to zero) to evaluate the loop integral. In cutoff regularization
one would get in addition a term linear in the cutoff and the density $\rho$. 
Since such a term is indistinguishable from the effect of a zero-range contact
interaction it does not belong to the genuine long-range contributions.}   
\begin{equation} U_\Sigma(k_f)^{(2\pi\Sigma)} = {F^2 g_A^2 M_B\,m_\pi^2 \over 
8\pi^3 f_\pi^4} \bigg\{- 3m_\pi  k_f+ ( 2k_f^2 +3m_\pi^2) \arctan{k_f \over 
m_\pi} \bigg\}\,, \end{equation}
with $m_\pi = 138\,$MeV the average pion mass. The mean baryon mass $M_B =
(2M_N+M_\Lambda+M_\Sigma)/4 = 1047\,$MeV serves the purpose of averaging 
out small differences in the kinetic energies of the various baryons involved.
Note the large scale enhancement factor $M_B$ in eq.(3) which stems from the 
energy denominator of the $2\pi$-exchange diagram. Because of this 
characteristic property the notion of iterated (second order) one-pion 
exchange is actually more appropriate (see also Sec. 4 in ref.\cite{nnpap} for
the analogous classification of the $2\pi$-exchange NN-interaction). The third
diagram in Fig.\,1 with two medium insertions represents a Pauli blocking
correction. With an intermediate $\Sigma$-hyperon the contribution to the real
part of the $\Sigma$-nuclear mean-field can be expressed as:
\begin{equation} U_\Sigma(k_f)^{(2\pi\Sigma)}_{\rm Pauli} = {F^2 g_A^2 M_B
\over (2\pi  f_\pi)^4}\bigg\{2k_f^4-m_\pi^4 \int_0^u dx -\!\!\!\!\!\!\int_{-1}
^1 {dz \over z}  \bigg[{1\over S} +2\ln S \bigg] \bigg\}\,, \end{equation} 
with the auxiliary function $S = 1+u-x+2x z^2+ 2z  \sqrt{x(u-x+x z^2)}$ and
the abbreviation $u= k_f^2/m_\pi^2$. The symbol $-\!\!\!\!\!\int_{-1}^1 dz$ 
in eq.(4) denotes a principal value integral. We also note that the total 
imaginary part $W_\Sigma(k_f)^{(2\pi\Sigma)}+W_\Sigma(k_f)^{(2\pi\Sigma)}_{\rm
Pauli}$ vanishes identically.    

Next, we come to the iterated one-pion exchange diagrams with an intermediate 
$\Lambda$-hyperon. The small $\Sigma\Lambda$-mass splitting $M_\Sigma-
M_\Lambda =77.5\,$MeV which comes here into play is comparable in magnitude to
the typical kinetic energies of the nucleons. Therefore it has to be counted 
accordingly in the energy denominator. By the relation $M_\Sigma-M_\Lambda = 
\Delta^2/M_B$ we introduce another small mass scale $\Delta$, whose magnitude 
$\Delta \simeq 285\,$MeV is close to the Fermi momentum $k_{f0}\simeq
263\,$MeV at nuclear matter saturation density. Putting all pieces together we
find from the second diagram in Fig.\,1 with an intermediate $\Lambda$-hyperon
the following long-range contribution to the complex $\Sigma$-nuclear mean 
field:  
\begin{equation} U_\Sigma(k_f)^{(2\pi\Lambda)}+i\,W_\Sigma(k_f)^{(2\pi\Lambda)}
={D^2 g_A^2 M_B\, m_\pi^4 \over 48\pi^3  f_\pi^4} \,\Psi\bigg({k_f^2 \over 
m_\pi^2},{\Delta^2 \over m_\pi^2}\bigg)\,, \end{equation}
where the complex function 
\begin{eqnarray}\Psi(u, \delta) &=& -(\delta+3)\sqrt{u} -{i \over 4}(u+2\delta
+6)\sqrt{u(4\delta+u)} + (2u+\delta^2+4\delta+3 )\nonumber \\ && 
\times \Bigg\{ \arctan{ \sqrt{u} \over 1+\delta} +i\, \ln { 2+2\delta +u + 
\sqrt{u(4\delta+u)} \over 2 [(1+\delta)^2+u]^{1/2}} \Bigg\}\,,\end{eqnarray}  
emerges from the combined loop and Fermi sphere integration with the
abbreviation $\delta= \Delta^2/m_\pi^2$. The corresponding Pauli blocking 
correction to the real part of the $\Sigma$-nuclear mean field can be 
expressed as a numerically easily manageable double-integral of the form: 
\begin{eqnarray}U_\Sigma(k_f)^{(2\pi\Lambda)}_{\rm Pauli} &=& {D^2g_A^2 M_B\,
m_\pi^4\over 6(2\pi  f_\pi)^4} \int_0^u dx \int_0^u dy \,{1\over (2 \delta+
1+x-y)^2 }\,\Bigg\{ {4 (y-x-2\delta-1) \sqrt{x y} \over (1+x+y)^2-4x y} 
 \\  &&+ (2x-2y +4\delta+1)\ln {1+x+y + 2\sqrt{x y}\over 1+x+y -2 
\sqrt{x y}}+ (2\delta +x -y)^2 \ln{|\delta -y-\sqrt{x y}|\over |\delta -y +
\sqrt{x y}|}\Bigg\}\,. \nonumber \end{eqnarray} 
In the case of the imaginary part of the $\Sigma$-nuclear mean field the Pauli
blocking correction can even be written in closed analytical form: 
\begin{eqnarray}W_\Sigma(k_f)^{(2\pi\Lambda)}_{\rm Pauli} &=& {D^2g_A^2 M_B\,
m_\pi^4\over 96\pi^3  f_\pi ^4} \, \theta(\sqrt{2} k_f - \Delta) \,\Bigg\{
{u^2 \over 4} +{3\over 2} (u- \delta^2) +(u-3)\delta \nonumber \\  &&+ {1\over
4}(u+2\delta+6) \sqrt{u(4\delta+u)} -(2u+\delta^2) \ln { 2+2\delta +u + 
\sqrt{u(4\delta+u)} \over 2+2u^{-1}\delta^2 }  \nonumber \\  && -(3+4
\delta) \ln { 2+2\delta +u + \sqrt{u(4\delta+u)} \over 2+4 \delta} \Bigg\} \,.
\end{eqnarray} 
It is interesting to observe that there is a threshold condition $k_f >\Delta/
\sqrt{2}$ for Pauli blocking to become active in the imaginary part of the 
$\Sigma$-nuclear mean field. This threshold corresponds to the subnuclear 
density $\rho_{\rm th} = 0.072\,{\rm fm}^{-3} = 0.45\,\rho_0$. Furthermore, 
as a simple check one verifies that the total imaginary $\Sigma$-nuclear mean 
field $W_\Sigma(k_f)^{(2\pi\Lambda)}+W_\Sigma(k_f)^{(2\pi\Lambda)}_{\rm Pauli}$
vanishes identically in the limit of $\Sigma \Lambda$-mass degeneracy, i.e.
$\delta = 0$. We have also evaluated the contributions to $U_\Sigma(k_f)$ from
irreducible two-pion exchange (which do not carry the large scale enhancement
factor $M_B$) and found that they sum up to zero in isospin-symmetric nuclear
matter. For comparison, the same exact cancellation is at work in the isoscalar
central channel of $2\pi$-exchange NN-potential (see Sec. 4.2 in
ref.\cite{nnpap}). As an aside we note that the Pauli blocking corrections
eqs.(4,7,8) could also be interpreted as the effects of the $2\pi$-exchange
$\Sigma NN$ three-body interaction \cite{usmani}. This equivalence becomes 
immediately clear by opening the two nucleon lines (with horizontal 
double-lines) of the last diagram in Fig.\,1.

\begin{figure}
\begin{center}
\includegraphics[scale=0.7]{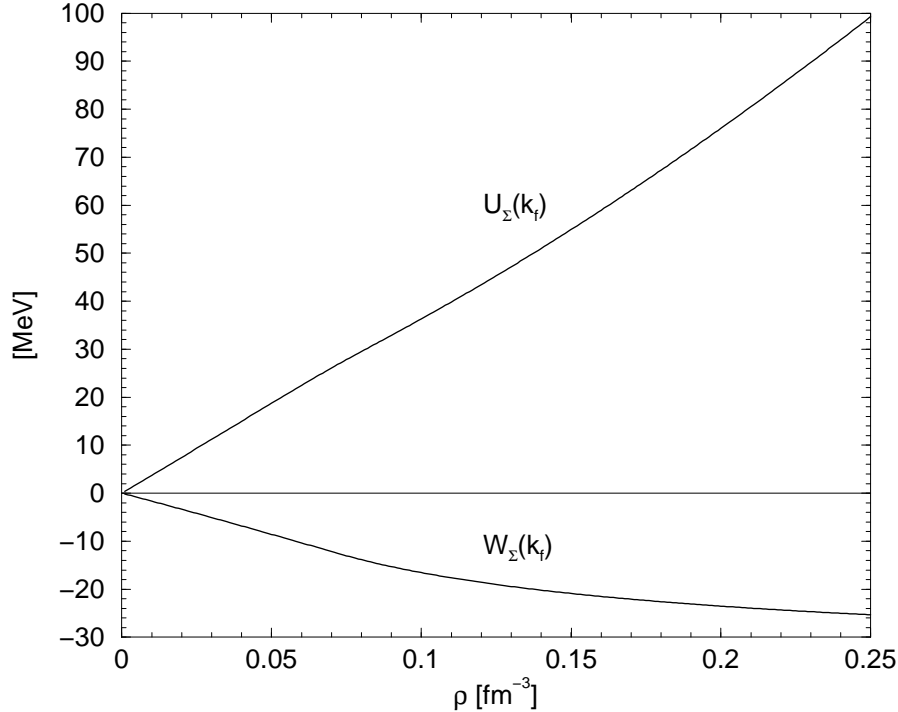}
\end{center}\vspace{-.8cm}
\caption{The complex mean field $U_\Sigma(k_f)+i\, W_\Sigma(k_f)$ of a
zero-momentum $\Sigma$-hyperon in isospin-symmetric nuclear matter versus the
nucleon density $\rho= 2k_f^3/3\pi^2$. The imaginary part $W_\Sigma(k_f)$
(lower solid line) originates from the conversion process $\Sigma N \to 
\Lambda N$ induced by one-pion exchange. The real part $U_\Sigma(k_f)$ (upper 
solid line) includes only genuine long-range contributions.}    
\end{figure}

Summing up all calculated terms, we show in Fig.\,2 the resulting complex 
$\Sigma$-nuclear mean field $U_\Sigma(k_f)+i\, W_\Sigma(k_f)$ as a function of
the nucleon density $\rho = 2k_f^3/3\pi^2$. The lower curve for the imaginary
part $W_\Sigma(k_f)$ displays clearly the onset of the Pauli blocking 
effects at the threshold density $\rho_{\rm th} = 0.072\,$fm$^{-3}$. It is very
astonishing that the total real $\Sigma$-nuclear mean field $U_\Sigma(k_f)$ 
follows to a good approximation a straight line, whereas individual components
possess a much more nonlinear dependence on the density $\rho$, driven by the 
relevant dimensionless variable $\sqrt{u} = k_f/m_\pi$. At normal nuclear 
matter density  $\rho_0 = 0.16\,$fm$^{-3}$ (corresponding to a Fermi momentum 
of $k_{f0}=263 \,$MeV) one finds for the real and imaginary part, $U_\Sigma(
k_{f0}) = [0.4+(40.9+16.1)+ (8.2-6.6)]\,$MeV $= 59\,$MeV and $W_\Sigma(k_{f0})
= (-29.0+7.5)\,$MeV $= -21.5\,$MeV, where the individual entries correspond to
the respective terms written in eqs.(2-8), in that order. As it could be 
expected from the small $KN\Sigma$-coupling constant $g_{KN\Sigma}= 4.4$ the
kaon-exchange contribution is completely negligible. The genuine long-range
terms from iterated one-pion exchange with intermediate $\Sigma$- and
$\Lambda$-hyperons eqs.(3,5) build up a sizeable repulsion of $49\,$MeV which
is furthermore enhanced (by about $20\%$) by the Pauli blocking corrections
eqs.(4,7). Note that one is dealing here with six-dimensional principal value
integrals whose signs are not a priori fixed. The imaginary single-particle
potential of $W_\Sigma(k_{f0})= -21.5\,$MeV comes out surprisingly close to
the value $-20\,$MeV obtained in the SU(6) quark model calculation of
ref.\cite{kohno2} or the value $-16\,$MeV extracted from $\Sigma^-$-atom data
\cite{dover}. Of course, in order to account for the uncertainties in the
axial-vector coupling constants $D$ and $F$ and the choice of a mean baryon
mass $M_B$ one should add to the curves in Fig.\,2 an error band of at least
$\pm 20\%$. Moreover, there are recoil corrections to the leading order
results given in eqs.(2-8). Since these recoil corrections scale (at least) as
$1/M_B$ with the baryon mass $M_B$ they are expected to suppressed by the
small relative factor $(k_f/M_B)^2 \leq 0.07$ (for moderate densities $\rho
\leq 0.2\,$fm$^{-3}$). If one continues the curves in Fig.\,2 to even higher
densities $\rho \leq 0.5\, $fm$^{-3}$ one finds a stronger than linear rise of
the real part $U_\Sigma(k_f)$ and an approximate saturation of the imaginary
part $U_\Sigma(k_f)$ at a value of about $-30\,$MeV. However, this behavior
should not be taken too seriously, since for Fermi momenta $k_f >350\,$MeV one
presumably exceeds the limits of validity of the present calculation based on
in-medium chiral perturbation theory (see also the discussion in
ref.\cite{nucmat}).

Altogether, it seems that the leading long-range two-pion exchange dynamics 
evaluated in the present field-theoretical approach is able to reproduce
qualitatively the single-particle properties of $\Sigma$-hyperons in the 
nuclear medium in agreement with existing phenomenology and more sophisticated 
model calculations. This raises the question about the role of the short-range 
$\Sigma N$-interaction (and additional short-range correlations). QCD sum rule 
calculations of $\Sigma$-hyperons in nuclear matter \cite{jin} indicate that
the individually large Lorentz scalar and vector mean fields (typically of
strength $\mp 0.2 M_\Sigma$) cancel each other to a large extent. In the case
of the Lorentz scalar mean field the QCD sum rule results are subject to large
uncertainties due to unknown contributions from four-quark condensates. 
Conversely, at least some of these contributions are accounted for by our
explicit treatment of the long-range two-pion exchange processes. A combined 
QCD sum rule analysis of nucleons, $\Lambda$-hyperons and $\Sigma$-hyperons in
nuclear matter together with input from in-medium chiral perturbation theory
and phenomenology should help to better constrain the short-distance
baryon-nucleon dynamics (whose details are however not resolved at the scale
of the nuclear Fermi momentum $k_{f0}$).  

In summary, we have calculated in this work the density-dependent complex 
mean field $U_\Sigma(k_f)+ i\, W_\Sigma(k_f)$ of a $\Sigma$-hyperon in
isospin-symmetric nuclear matter in the two-loop approximation of in-medium 
chiral perturbation theory. The leading long-range $\Sigma N$-interaction 
arises from iterated (second order) one-pion exchange with a $\Sigma$- or 
$\Lambda$-hyperon in the intermediate state. These second order pion-exchange 
contributions do not correspond to any mean field Hartree approximation as 
evidenced by their intrinsic nonlinear density-dependence. To the order in the
small momentum expansion we are working here the long-range correlations 
between the $\Sigma$-hyperon and the nucleons are fully taken into account. We
find from the strong $\Sigma N\to \Lambda N$ conversion process at nuclear
matter saturation density $\rho_0 = 0.16\,$fm$^{-3}$ an imaginary
single-particle potential of $W_\Sigma(k_{f0}) =-21.5$\,MeV. The genuine
long-range contributions from iterated pion-exchange sum up to a moderately
repulsive real single-particle potential of $U_\Sigma(k_{f0})=59$\,MeV. Taking
into account the uncertainties of the involved coupling constants such values
are already compatible with the $\Sigma$-nucleus optical potential needed to 
describe the inclusive $(\pi^-,K^+$) spectra related to $\Sigma^-$-formation 
in heavy nuclei. Our results suggest that the net effect of the short-range 
$\Sigma N$-interaction on the $\Sigma$-nuclear mean-field could be small. A 
combined QCD sum rule analysis of nucleons,  $\Lambda$-hyperons and 
$\Sigma$-hyperons in nuclear matter can help to clarify the latter
point. Furthermore, the present calculation can easily be extended to nonzero 
$\Sigma$-momentum $p$, and from the corresponding momentum and density 
dependent mean field $U_\Sigma(p,k_f)+ i\,W_\Sigma(p,k_f)$ one can extract, for
example, an effective $\Sigma$-mass \cite{usmani,mill}. Work along these lines 
is in progress.            

\bigskip

Acknowledgements: I thank A. Gal, G. Lalazissis and W. Weise for informative 
discussions.

\end{document}